\documentclass[twoside]{elsart}
\usepackage{epsfig,multirow}
\usepackage{array}
\usepackage{amssymb,amsbsy,amsmath}
\usepackage{times}
\newcommand{\figref}[1]{Fig.~\protect\ref{#1}}
\newcommand{\secref}[1]{Sec.~\protect\ref{#1}}
\renewcommand{\eqref}[1]{Eq.~(\protect\ref{#1})}
\newcommand{\xref}[1]{\protect\ref{#1}}
\newcommand{\fmref}[1]{(\protect\ref{#1})}

\newcommand{\element}[2]{$^{#1}$#2}
\def\V0{\stackrel{\circ}{V}}
\def\v0{\stackrel{\circ}{v}}
\def\half{{\frac{1}{2}}}

\newcommand{\op}[1]{%
    \fontdimen12\textfont3=2pt\fontdimen12\scriptfont3=1.4pt%
    \!\null\mathop{\vphantom{#1}\smash{#1}}\limits_{\sim}\null\!}

\def\bra#1{\langle \, {#1} \, | \,}
\def\ket#1{\, | \, {#1} \, \rangle}
\newcommand{\braket}[2]{\langle \, {#1} \, | \, {#2} \, \rangle}

\def\ketsm{\ket{s\,m}}
\def\brasm{\bra{s\,m}}

\def\ketsm1{\ket{s_1\,m_1}}
\def\brasm1{\bra{s_1\,m_1}}
\def\ketsm2{\ket{s_2\,m_2}}
\def\brasm2{\bra{s_2\,m_2}}
\newcommand{\dd}[2]{\frac{{d}\, {#1}}{{d} {#2}}\;}

\newcommand{\vek}[1]{{\!\vec{\,#1}}}

\newcommand{\SMax}{S_{\mbox{\scriptsize max}}}
\newcommand{\SMin}{S_{\mbox{\scriptsize min}}}
\begin{document}
%---------- Titel und Abstract-------------------------------
%
\typeout{   --- >>>   relevant dimension of the Heisenberg model   <<<   ---   }
\typeout{   --- >>>   relevant dimension of the Heisenberg model   <<<   ---   }
\typeout{   --- >>>   relevant dimension of the Heisenberg model   <<<   ---   }
%
%---------- Journal-----------------------------------------
%
\journal{Journal of Magnetism and Magnetic Materials}
\begin{frontmatter}
\title{Structure and relevant dimension of the Heisenberg model
and applications to spin rings}
 
%\author{K. B\"arwinkel, H.-J. Schmidt, J. Schnack}
\author{K. B\"arwinkel, H.-J. Schmidt, J. Schnack\thanksref{JS}}
\address{Universit\"at Osnabr\"uck, Fachbereich Physik \\ 
         Barbarastr. 7, 49069 Osnabr\"uck, Germany}
\thanks[JS]{corresponding author: jschnack\char'100uos.de,\\ http://www.physik.uni-osnabrueck.de/makrosysteme/}

\begin{abstract}
\noindent
For the diagonalization of the Hamilton matrix in the Heisenberg
model relevant dimensions are determined depending on the
applicable symmetries. Results are presented, both, by general
formulae in closed form and by the respective numbers for a
variety of special systems. In the case of cyclic symmetry,
diagonalizations for Heisenberg spin rings are performed with
the use of so-called magnon states. Analytically solvable cases
of small spin rings are singled out and evaluated.

\vspace{1ex}

\noindent{\it PACS:} % \\
67.57.Lm;          % Spin dynamics
75.10.Jm;          % Quantized spin models \\
75.40.Cx;          % Static properties of magnetic systems \\
75.40.Gb           % Dynamic properties of magnetic systems \\
\vspace{1ex}

\noindent{\it Keywords:} Heisenberg model; spin ring; magnon
states; diagonalization; relevant dimension
\end{abstract}
\end{frontmatter}
%%%%%%%%%%%%%%%%%%%%%%%%%%%%%%%%%%%%%%%%%%%%%%%%%%%%%%%%%%%%%%%%%%%%%%%%
%\newpage
%\tableofcontents
%%%%%%%%%%%%%%%%%%%%%%%%%%%%%%%%%%%%%%%%%%%%%%%%%%%%%%%%%%%%%%%%%%%%%%%%
%\newpage
\raggedbottom
\section{Introduction and summary}

As far as their magnetic behaviour is concerned, some recently
synthesized mole\-cules like ``ferric wheels" of six, eight or
ten iron ions of spin $\frac{5}{2}$
\cite{TDP94,GCR94,FST96,LGB97} appear as a limited array of
localized single-particle spins which are adequately described
by the Heisenberg model \cite{BeG90,DGP93,CCF96,PDK97}.  The
calculation of key quantities like for example the spin-spin
correlation function becomes easy once the Heisenberg
Hamiltonian has been diagonalized. For a straight forward
diagonalization the dimension $d$ of the Hilbert space
${\mathcal H}$, which for instance is $d=(2\,s+1)^N$ for $N$
spins with spin quantum number $s$, may, however, appear
prohibitively large even for rather small systems. But the
obvious symmetries allow to reduce the problem to a set of
less ample problems according to a decomposition of ${\mathcal H}$
into a set of mutually orthogonal subspaces. 

In the case of spin arrays the following symmetry operators may
be employed, depending on the degree of symmetry exhibited by
the system, i.e. the Hamilton operator under consideration:
\begin{itemize}
\item the 3-component of the total spin in the case of axial symmetry,
\item the total spin in the case of full
rotational symmetry,
\item the cyclic shift operator for rings of identical single
spins with translationally invariant coupling,
\item further discrete symmetries like reflection of the spin
ring, reflection of the spin orientation and reflection of time order.
\end{itemize}
Group theoretical arguments are already used to calculate the
spectra of some molecules by means of an irreducible tensor
operator approach \cite{DGP93,WSK99}. In the present article an
alternative technique is devised. Also, general expressions for
the relevant dimension are derived without taking recourse to any
specific procedure of diagonalization.

We will generalize considerations by D.~Kouzoudis
\cite{Kou97,Kou98} who found analytical solutions for special
cases of small Heisenberg rings.  To this end we are
concentrating on the first three symmetries and denote the
corresponding simultaneous eigenspaces by ${\mathcal H}(S,M,p)$.
After introducing our notation in \secref{sec-2}, we will
investigate the various symmetries and their corresponding
subspaces in \secref{sec-3}. The dimension of the ${\mathcal
H}(S,M,p)$ can be calculated exactly and we derive an explicit
formula for the dimension of the subspaces ${\mathcal H}(S,M)$
and ${\mathcal H}(S,M,p)$, which enables the evaluation of the
relevant dimension $d_r$. In \secref{sec-4} we provide recursion
formulae to calculate the Hamilton matrix restricted to the
subspaces ${\mathcal H}(S,M,p)$. To this end tools from solid
state physics, namely $l$-magnon states, are utilized
\cite{Yos96,HoP40,Roe77}.  Section \xref{sec-5} is devoted to
examples and applications. For rings of $N=5, s=1$ and $N=8,
s=\half$ an exact diagonalization of the Heisenberg Hamiltonian
with nearest neighbour interaction has been performed using
Mathematica$^{\circledR}$ \cite{HJS99}. Further we compare the
analytical anti-ferromagnetic ground state energies for $N=2\dots 8$
with the asymptotic Bethe-Hulth\'{e}n formula for
$N\rightarrow\infty$ \cite{Bet31,Hul38}.

\section{Heisenberg model}
\label{sec-2}

The most general Hamilton operator considered in these
investigations consists of a spin-spin interaction term
$\op{H}_0$ and a term $\op{H}_F$ describing the interaction with an
external field which is assumed to point in the $3$-direction
%--------------------------------------------------------
\begin{eqnarray}
\label{E-2-1}
\op{H}
&=&
\op{H}_0 + \op{H}_F
=
-
\sum_{x,y}^N\;
J(x,y)\;
\op{\vek{s}}(x) \cdot \op{\vek{s}}(y)
-
\sum_{x}^N\;
\mu\, B\, \op{s}^3(x)
\ .
\end{eqnarray}
%--------------------------------------------------------
$x,y \in G=\{1,\dots,N\}$ label the spin sites modulo $N$ and
the upper right index denotes the spin component. All spins are
defined to be dimensionless vector observables with the
commutation relations
%--------------------------------------------------------
\begin{eqnarray}
\label{E-2-2}
\left[
\op{s}^a(x),\op{s}^b(y)
\right]
&=&
i\,\epsilon_{abc}\,\op{s}^c(x)\,\delta_{xy}
\ .
\end{eqnarray}
%--------------------------------------------------------
The eigenvalue of $(\op{\vek{s}}(x))^2$ is $s(x)(s(x)+1)$,
$s(x)$ being the individually fixed quantum number at site
$x$. The possible eigenvalues of any $\op{s}^3(x)$ are the
magnetic quantum numbers $m_x$. One may introduce the ladder
operators 
%--------------------------------------------------------
\begin{eqnarray}
\label{E-2-3}
\op{s}^{\pm}(x)
=
\op{s}^{1}(x) \pm i\,\op{s}^{2}(x)
\end{eqnarray}
%--------------------------------------------------------
to reformulate $\op{H}_0$ as
%--------------------------------------------------------
\begin{eqnarray}
\label{E-2-4}
\op{H}_0
&=&
-
\sum_{x,y\in G}\;
J(x,y)\,
\left\{
\op{s}^3(x) \op{s}^3(y)
+
\frac{1}{2}
\left[
\op{s}^+(x) \op{s}^-(y)
+
\op{s}^-(x) \op{s}^+(y)
\right]
\right\}
\ .
\end{eqnarray}
%--------------------------------------------------------
The total Hilbert space is spanned by the product basis of
the single-particle eigenstates of all $\op{s}^3(x)$
%--------------------------------------------------------
\begin{eqnarray}
\label{E-2-5}
\op{s}^3(x)\,
\ket{m_1, \dots, m_x, \dots, m_N}
=
m_x\,
\ket{m_1, \dots, m_x, \dots, m_N}
\ ,
\end{eqnarray}
%--------------------------------------------------------
which will be a reference basis in our investigations.

\section{Symmetry operations and relevant dimensions}
\label{sec-3}

Our most general Hamiltonian \fmref{E-2-1} allows only for
invariance with respect to rotations about the
$3$-axis. Obviously, $\op{H}_0$ as well as $\op{H}_F$ commute
with the $3$-component $\op{S}^3$ of the total spin
%--------------------------------------------------------
\begin{eqnarray}
\label{E-3-1}
\left[
\op{H},\op{S}^3
\right]
&=&
0
\ ,\quad \op{S}^3 = \sum_{x\in G}\;\op{s}^3(x)
\ ,
\end{eqnarray}
%--------------------------------------------------------
the possible eigenvalues of which are the total magnetic quantum
numbers
%--------------------------------------------------------
\begin{eqnarray}
\label{E-3-1A}
M=-\SMax, -\SMax+1, \dots, \SMax
\quad \mbox{with}\quad
\SMax
=
\sum_{x=1}^N\;s(x)
\end{eqnarray}
%--------------------------------------------------------
being the maximum total spin quantum number.
The total Hilbert space ${\mathcal H}$ for the Heisenberg model
is the direct sum of all eigenspaces ${\mathcal H}(M)$ of
$\op{S}^3$
%--------------------------------------------------------
\begin{eqnarray}
\label{E-3-2}
{\mathcal H}
=
\bigoplus_{M=-\SMax}^{+\SMax}\;
{\mathcal H}(M)
\ .
\end{eqnarray}
%--------------------------------------------------------
The problem of diagonalizing $\op{H}$ in ${\mathcal H}$ is thus
broken up into the corresponding problems in each of the
${\mathcal H}(M)$. This reduces the dimension one has to cope
with from
%--------------------------------------------------------
\begin{eqnarray}
\label{E-3-3}
d
=
\mbox{dim}\left({\mathcal H}\right)
=
\prod_{x=1}^N\;\left(2 s(x) + 1\right)
\end{eqnarray}
%--------------------------------------------------------
to the respective $\mbox{dim}\left({\mathcal H}(M)\right)$. The
relevant dimension $d_r$ then is the maximum of those
dimensions, if no other symmetries can be exploited.  For given
values of $M$, $N$ and of all $s(x)$ the dimension
$\mbox{dim}\left({\mathcal H}(M)\right)$ can be determined as
the number of product states \fmref{E-2-5}, which constitute a
basis in ${\mathcal H}(M)$, with $\sum_{x\in G}m_x=M$.  The
solution of this combinatorial problem can be given in closed
form.

{\bf Theorem:} 
%--------------------------------------------------------
\begin{eqnarray}
\label{E-3-5}
\mbox{dim}\left({\mathcal H}(M)\right)
=
\frac{1}{(\SMax-M)!}\, 
\left[
\left(\dd{}{z}\right)^{\SMax-M}\;
\prod_{x=1}^{N}\;
\frac{1-z^{2 s(x) + 1}}{1-z}
\right]_{z=0}
\ .
\end{eqnarray}
%--------------------------------------------------------
For equal single-spin quantum numbers
$s(1)=\cdots=s(N)=s$, and thus a maximum total spin quantum
number of $\SMax=Ns$, \eqref{E-3-5} simplifies to
%--------------------------------------------------------
\begin{eqnarray}
\label{E-3-6}
\mbox{dim}\left({\mathcal H}(M)\right)
&=&
f(N,2s+1,\SMax-M)
\qquad\mbox{with} \\
\label{E-3-6-A}
f(N,\mu,\nu)
&=&
\sum_{n=0}^{\left[\nu/\mu\right]}
(-1)^n\, \binom{N}{n}
\binom{N-1+\nu - n \mu}{N-1}
\ .
\end{eqnarray}
%--------------------------------------------------------
In both formulae \fmref{E-3-5} and \fmref{E-3-6}, $M$ may be
replaced by $|M|$ since 
the dimension of ${\mathcal H}(M)$ equals those of ${\mathcal
H}(-M)$. $\left[\nu/\mu\right]$ in the sum symbolizes the
greatest integer less or equal to $\nu/\mu$. 
\eqref{E-3-6} is known as a result of De Moivre \cite{Fel68}. 

{\bf Proof:} The proof of \fmref{E-3-5} and \fmref{E-3-6} may be
accomplished by comparing any product state \fmref{E-2-5} with
the completely aligned state
%--------------------------------------------------------
\begin{eqnarray}
\label{E-3-7}
\ket{\Omega} &=& \ket{m_1=s(1), m_2=s(2),\dots, m_N=s(N)}
\ ,
\end{eqnarray}
%--------------------------------------------------------
which is also called magnon vacuum state, see next
section. Evidently it is also an eigenstate of $\op{H}$ and in
the ferromagnetic case a ground state of $\op{H}_0$. 
Using the possible decrements $a(x)=s(x)-m_x$ of the magnetic quantum
number at each spin site we want to construct a generating function $F(z)$ of the
numbers $\mbox{dim}\left({\mathcal H}(M)\right)$ as a polynomial in $z$
%--------------------------------------------------------
\begin{eqnarray}
\label{E-3-9}
F(z)
=
\sum_{n=0}^{2 \SMax}\,
\mbox{dim}\left({\mathcal H}(\SMax-n)\right)
z^{n}
\ .
\end{eqnarray}
%--------------------------------------------------------
The consideration, that the number of product states with fixed
magnetic quantum number $M$ corresponds to the related number of
sequences of decrements, leads to the following definition
%--------------------------------------------------------
\begin{eqnarray}
\label{E-3-8}
F(z)
=
\prod_{x=1}^N
\left(
\sum_{a(x)=0}^{2 s(x)}\,
z^{a(x)}
\right)
=
\prod_{x=1}^N\,
\frac{1-z^{2 s(x) + 1}}{1-z}
\ .
\end{eqnarray}
%--------------------------------------------------------
Eq. \fmref{E-3-5} is then obvious and \fmref{E-3-6} is easily
inferred from \fmref{E-3-5} by using Leibniz's theorem
\cite{Abr73} to calculate multiple derivatives of the product
$(1-z^{2 s+1})^N\cdot (1-z)^{-N}$.  \hfill $\blacksquare$

In the special case of identical
single-particle spins, $\op{H}_F$ is proportional to $\op{S}^3$
and therefore commutes with $\op{H}_0$. The simultaneous
eigenstates of $\op{H}_0$ and $\op{S}^3$ are then also
eigenstates of $\op{H}_F$. $\op{H}_F$ causes a splitting
proportional to $M$ of the eigenvalues of $\op{H}_0$, which
otherwise would not depend on $M$.

In what follows, consideration is alternatively restricted to one
of two special cases. In the first case (I) the Hamilton
operator is field free, $\op{H}=\op{H}_0$, and thus invariant
under any rotation. In the second case (II) all individual spins
are identical.

The total spin quantum number will be denoted by $S$. It has
values in the set $\{0\le \SMin,\SMin+1,\dots,\SMax \}$ and 
$\op{\vek{S}}^2$ has eigenvalues $S(S+1)$. The simultaneous
eigenspaces ${\mathcal H}(S,M)$ of $\op{\vek{S}}^2$ and
$\op{S}^3$ are spanned by eigenvectors of $\op{H}$.
The one-dimensional subspace ${\mathcal H}(M=\SMax)={\mathcal
H}(\SMax,\SMax)$, especially, is spanned by $\ket{\Omega}$.
The total ladder operators are
%--------------------------------------------------------
\begin{eqnarray}
\label{E-3-11}
\op{S}^{\pm}
=
\op{S}^{1}
\pm
i\,\op{S}^{2}
\ .
\end{eqnarray}
%--------------------------------------------------------
For $S>M$,  $\op{S}^{-}$ maps any normalized
$\op{H}$-eigenstate 
$\in{\mathcal H}(S,M+1)$ onto an $\op{H}$-eigenstate
$\in{\mathcal H}(S,M)$ with norm $\sqrt{S(S+1)-M(M+1)}$. 

{\bf Lemma:}
For $0 \le M < \SMax$, ${\mathcal H}(M)$ can be decomposed
into orthogonal subspaces
%--------------------------------------------------------
\begin{eqnarray}
\label{E-3-12}
{\mathcal H}(M)
=
{\mathcal H}(M,M)
\oplus
\op{S}^{-}{\mathcal H}(M+1)
\end{eqnarray}
%--------------------------------------------------------
with
%--------------------------------------------------------
\begin{eqnarray}
\label{E-3-13}
\op{S}^{-}{\mathcal H}(M+1)
=
\bigoplus_{S\ge M+1}
{\mathcal H}(S,M)
\ .
\end{eqnarray}
%--------------------------------------------------------
{\bf Proof:} The ${\mathcal H}(S,M)$ represent an orthogonal
decomposition of ${\mathcal H}(M)$. It is therefore sufficient
to identify the orthogonal complement of $\op{S}^{-}{\mathcal
H}(M+1)$ within ${\mathcal H}(M)$ as ${\mathcal H}(M,M)$. This,
in turn, is clear if any $\ket{\psi}\neq 0 \in {\mathcal H}(M)$
with $\ket{\psi}\perp\op{S}^{-}{\mathcal H}(M+1)$ vanishes on
application of $\op{S}^{+}$. The defining property of all such
$\ket{\psi}$ is
%--------------------------------------------------------
\begin{eqnarray}
\label{E-3-14}
\bra{\psi}\op{S}^{-}\ket{\phi}=0
\quad \forall \ket{\phi}\in {\mathcal H}(M+1)
\ .
\end{eqnarray}
%--------------------------------------------------------
But then $\op{S}^{+}\ket{\psi}\in {\mathcal H}(M+1)$ and
$\braket{\op{S}^{+}\psi}{\phi}=0$ which proves
$\op{S}^{+}\ket{\psi}=0$.\hfill $\blacksquare$

In consequence, the diagonalization of $\op{H}$ in ${\mathcal
H}$ has now been traced back to diagonalization in the subspaces 
${\mathcal H}(S,S)$, the dimension of which are for $S<\SMax$
%--------------------------------------------------------
\begin{eqnarray}
\label{E-3-15}
\mbox{dim}\left({\mathcal H}(S,S)\right)
=
\mbox{dim}\left({\mathcal H}(M=S)\right)
-
\mbox{dim}\left({\mathcal H}(M=S+1)\right)
\end{eqnarray}
%--------------------------------------------------------
and can be calculated according to \eqref{E-3-5} in case (I) or
\eqref{E-3-6} in case (II). The relevant dimension $d_r$ is the
maximum of the numbers given by \eqref{E-3-15}.

For many of the systems of interest here, the individual spin
quantum numbers are equal, the coupling coefficients $J$ have
the special property $J(x,y)=J(|x-y|)$ and the applied magnetic
field is homogeneous. This renders the Hamiltonian invariant
with respect to translations on the group $G=\{1,\dots,N\}$ of
spin sites . Any such translation is represented by the cyclic shift
operator $\op{T}$ or a multiple repetition. $\op{T}$ is defined
by its action on the product basis \fmref{E-2-5}
%--------------------------------------------------------
\begin{eqnarray}
\label{E-3-16}
\op{T}\,
\ket{m_1, \dots, m_{N-1}, m_N}
=
\ket{m_N, m_1, \dots, m_{N-1}}
\ .
\end{eqnarray}
%--------------------------------------------------------
The eigenvalues of $\op{T}$ are the $N$-th roots of unity,
$z=\exp\{-i p \}$, where $p$ will be called magnon momentum and
can take the following $N$ values modulo $2\pi$ from
%--------------------------------------------------------
\begin{eqnarray}
\label{E-3-17}
\widehat{G}
=
\left\{
p\ |\ 
p=\frac{2\pi k}{N},\ k=0,\dots, N-1
\right\}
\ ,
\end{eqnarray}
%--------------------------------------------------------
where $k$ will be called translational quantum number.
Clearly, $\op{T}$ commutes now with both Hamiltonian and total
spin. Any ${\mathcal H}(S,M)$ is decomposed into simultaneous
eigenspaces ${\mathcal H}(S,M,p)$ of $\op{\vek{S}}^2$,
$\op{S}^3$ and $\op{T}$, and diagonalization can be reduced to a
diagonalization in the subspaces ${\mathcal H}(S,S,p)$.
The according reduction of
the relevant dimension can be quantified if the values of $p$
and their degeneracy within ${\mathcal H}(S,S)$ are known. As a
rule of thumb one finds a tendency for equal degeneracy and thus
%--------------------------------------------------------
\begin{eqnarray}
\label{E-3-18}
d_r
\approx
\frac{1}{N}\,
\max_{S} \mbox{dim}\left({\mathcal H}(S,S)\right)
\ .
\end{eqnarray}
%--------------------------------------------------------
It is, however, possible to determine $d_r$ exactly for any $N$
and $s$. To this end we introduce the notations of cycles,
proper cycles and epicycles. A special decomposition of
${\mathcal H}$ into orthogonal subspaces can be achieved by
starting with the product basis \fmref{E-2-5} and considering
the equivalence relation
%--------------------------------------------------------
\begin{eqnarray}
\label{E-3-19}
\ket{\psi}\cong\ket{\phi}
\Leftrightarrow
\ket{\psi}=\op{T}^n\,\ket{\phi}
\ ,\ n\in\{ 1,2,\dots,N\}
\end{eqnarray}
%--------------------------------------------------------
for any pair of states belonging to the product basis. The
equivalence relation then induces a complete decomposition of
the basis into disjoint subsets, i.e. the equivalence classes.
A ``cycle" is defined as the linear span of such an equivalence
class of basis vectors. The obviously orthogonal
decomposition of ${\mathcal H}$ into cycles is compatible with
the decomposition of ${\mathcal H}$ into the various ${\mathcal
H}(M)$ but not, generally, with the decomposition of ${\mathcal
H}(M)$ into its subspaces ${\mathcal H}(S,M)$. Evidently, the
dimension of a cycle can never exceed $N$. Cycles are called
``proper cycles" if their dimension equals $N$, they are termed
``epicycles" else. One of the $N$ primary basis states of a
proper cycle may arbitrarily be denoted as
%--------------------------------------------------------
\begin{eqnarray}
\label{E-3-20}
\ket{\psi_1}
=
\ket{m_1, \dots, m_x, \dots, m_N}
\end{eqnarray}
%--------------------------------------------------------
and the remaining ones may be enumerated as 
%--------------------------------------------------------
\begin{eqnarray}
\label{E-3-21}
\ket{\psi_{n+1}}
=
\op{T}^n\,\ket{\psi_1}
\ ,\ n=1,2,\dots,N-1
\ .
\end{eqnarray}
%--------------------------------------------------------
The cycle under consideration is likewise spanned by the states 
%--------------------------------------------------------
\begin{eqnarray}
\label{E-3-22}
\ket{\chi_k}
=
\frac{1}{\sqrt{N}}
\sum_{\nu=0}^{N-1}\,
\left(
e^{i\frac{2\pi\,k}{N}}
\op{T}
\right)^{\nu}
\ket{\psi_{\nu+1}}
\end{eqnarray}
%--------------------------------------------------------
which are eigenstates of $\op{T}$ with the respective magnon
momentum $p(k)=2\pi k/N$. Consequently, every $k$ (every $p(k)$)
occurs once in a proper cycle.

The reader will easily verify the validity of the following
related statements: 
\begin{itemize}
\item An epicycle of dimension $D$ is spanned by $D$ eigenstates
of $\op{T}$ with each of the translational quantum numbers
$k=0,N/D,\dots,(D-1)N/D$ occurring exactly once. 
\item For a primary product state in an epicycle of dimension
$D$, the spin sites are grouped into $D$ ``subrings", each
subring having a constant magnetic decrement $s-m_x=:a$ on all
of its places. The subring forms a pattern which repeats itself on
a cyclic shift of $D$ steps. The total number of spin sites in a
subring is $N/D$.
\item Therefore, for $M<\SMax$, ${\mathcal H}(M)$ contains a
cycle (epi or proper) of dimension $D$ if and only if $D$ is a
divisor of $N$, including $D=1$ and $D=N$, and $D(\SMax-M)/N$
is an integer. The set of such allowed $D$
will be denoted as $C(N,s,M)$.
\end{itemize}
Let ${\mathcal K}_M(D)$ be the linear span of all cycles of
dimension $D$ occurring within ${\mathcal H}(M)$ and let
$k_s(N,M,D)$ denote its dimension. According to the above
remarks we have
%--------------------------------------------------------
\begin{eqnarray}
\label{E-3-23}
\hspace*{-5mm}
\mbox{dim}\left({\mathcal H}(M)\right)
&=&
\sum_{D\in C(N,s,M)}\, k_s(N,M,D)
\\
&=&
k_s(N,M,N) + \sum_{D\in C(N,s,M);D\neq N}\, k_s(N,M,D)
\nonumber
\ .
\end{eqnarray}
%--------------------------------------------------------
Because to
each epicycle of dimension $D$ there corresponds exactly one
proper cycle with $D$ spin sites and total magnetic quantum number $M
D / N$, and vice versa,
\fmref{E-3-23} may be rewritten as 
%--------------------------------------------------------
\begin{eqnarray}
\label{E-3-24}
\mbox{dim}\left({\mathcal H}(M)\right)
&=&
k_s(N,M,N) + \sum_{D\in C(N,s,M)}\, k_s(D,\frac{M D}{N},D)
\ ,
\end{eqnarray}
%--------------------------------------------------------
which, together with \fmref{E-3-6}, may be used as a recursion
relation for the function $k_s(N,M,N)$ and hence also for
$k_s(N,M,D)$. 
This recursion relation can be transformed into an explicit
formula which reads
%--------------------------------------------------------
\begin{eqnarray}
\label{E-3-27}
k_s(N,M,N)
=
\sum_{D\in C(N,s,M)}\, 
q\left(\frac{N}{D} \right)\,
f\left(D,2s+1,\frac{D(\SMax-M)}{N} \right)
\end{eqnarray}
%--------------------------------------------------------
where $f$ is taken over from \eqref{E-3-6-A} and
%--------------------------------------------------------
\begin{eqnarray}
\label{E-3-28}
q\left(\nu\right)
=
\begin{cases}
(-1)^m  & \text{if $\nu$ is a product of $m$ different primes,}\\
0       & \text{else.}
\end{cases}
\end{eqnarray}
%--------------------------------------------------------
The proof rests on some elementary combinatorics and will be
omitted here. Further we have for the simultaneous eigenspaces
${\mathcal H}(\cdot,M,p(k))$ of $\op{S}^3$ and $\op{T}$
%--------------------------------------------------------
\begin{eqnarray}
\label{E-3-25}
\mbox{dim}\left({\mathcal H}(\cdot,M,p(k))\right)
&=&
\sum_{\stackrel{\scriptstyle u=1}{u\,|N;u\,|k}}^N\, k_s\left(N,M,\frac{N}{u}\right)
\frac{u}{N}
\end{eqnarray}
%--------------------------------------------------------
and 
%--------------------------------------------------------
\begin{eqnarray}
\label{E-3-26}
\mbox{dim}\left({\mathcal H}(S,M,p)\right)
=\mbox{dim}\left({\mathcal H}(\cdot,S,p)\right)-
\mbox{dim}\left({\mathcal H}(\cdot,S+1,p)\right)
\end{eqnarray}
%--------------------------------------------------------
for $0\le S < \SMax$ and $|M|\le S$.
This allows the explicit calculation of the relevant dimensions
for any given $N$ and $s$, for examples see table \xref{T-3-1}.

\section{Diagonalization of the Hamiltonian in the magnon basis representation}
\label{sec-4}

In this section we consider the Hamiltonian \fmref{E-2-1} with
$J(x,y)=J(|x-y|)$, $B=0$ and equal single spin quantum
numbers
%--------------------------------------------------------
\begin{eqnarray}
\label{E-4-1}
\op{H}
&=&
-
\sum_{x,y\in G}\;
J(|x-y|)\;
\op{\vek{s}}(x) \cdot \op{\vek{s}}(y)
\ ,\quad \forall x\in G: s(x)=s
\ .
\end{eqnarray}
%--------------------------------------------------------
In order to calculate the matrix elements of this Hamiltonian
restricted to subspaces ${\mathcal H}(S,M,p)$ it is
recommendable to use a basis of vectors which are already
adapted to the problem. We found it most convenient to work with
a basis constructed from so-called magnon states, used for
example in \cite{Roe77}. These magnon states should not be
confused with those defined by Holstein and Primakoff
\cite{HoP40}. The pertinent definitions are the following.

For any function $f$ of spin sites $x\in G$ the discrete Fourier
transform is as usual defined by
%--------------------------------------------------------
\begin{eqnarray}
\label{E-4-2}
f_p
:=
\frac{1}{\sqrt{N}}
\sum_{x\in G}\; e^{i p x}\, f(x)
\ ,\ p\in\widehat{G}\ .
\end{eqnarray}
%--------------------------------------------------------
The analogous transformation may be applied to linear operators
$\op{A}(x)$, where $\op{A}(x)$ for instance may be a
single-particle spin component at site $x$
%--------------------------------------------------------
\begin{eqnarray}
\label{E-4-4}
\op{S}_p^j
:=
\frac{1}{\sqrt{N}}
\sum_{x\in G}\; e^{i p x}\, \op{s}^j(x)
\end{eqnarray}
%--------------------------------------------------------
or a ladder operator \fmref{E-2-3}, which yields
%--------------------------------------------------------
\begin{eqnarray}
\label{E-4-5}
\op{S}_p^{\pm}
=
\op{S}_p^{1}
\pm
i\,\op{S}_p^{2}
\ .
\end{eqnarray}
%--------------------------------------------------------
Then the Hamiltonian \fmref{E-4-1} can be written in terms of the
$\op{S}_p^{\pm}$ and $\op{S}_p^{3}$ as (see \cite{Roe77})
%--------------------------------------------------------
\begin{eqnarray}
\label{E-4-6}
\op{H}
&=&
-
\sqrt{N}\sum_{p\in\widehat{G}}\;
J_p\,\op{\vek{S}}_p^* \cdot \op{\vek{S}}_p
\\
&=&
-
\sum_{p\in\widehat{G}}\;
\tilde{J}_p\,
\left(
\op{S}_{-p}^{+}\op{S}_{p}^{-} + \op{S}_{-p}^{3}\op{S}_{p}^{3}
\right)
\ ,
\quad
\tilde{J}_p:=\sqrt{N}J_p = \sum_{x\in G}\; e^{i p x}\, J(x)
\nonumber
\ .
\end{eqnarray}
%--------------------------------------------------------
It is straight forward to obtain the following commutation
relations
%--------------------------------------------------------
\begin{eqnarray}
\label{E-4-7-1}
\left[ \op{S}_{p}^{+}, \op{S}_{q}^{+} \right]
&=&
\left[ \op{S}_{p}^{-}, \op{S}_{q}^{-} \right]
=
0\ ,
\\
\label{E-4-7-2}
\left[ \op{S}_{p}^{+}, \op{S}_{q}^{-} \right]
&=&
\frac{2}{\sqrt{N}}\op{S}_{p+q}^{3}\ ,
\qquad
\left[ \op{S}_{p}^{3}, \op{S}_{q}^{\pm} \right]
=
\pm\frac{1}{\sqrt{N}}\op{S}_{p+q}^{\pm}
\end{eqnarray}
%--------------------------------------------------------
from the corresponding commutation relations of the spin
operators at site $x$. 

We define ``$l$-magnon states" with momenta
$p_1,\dots,p_l$ as
%--------------------------------------------------------
\begin{eqnarray}
\label{E-4-9}
\ket{\Omega_{\vek{p}}}
=
\ket{\Omega_{p_1,\dots,p_l}}
:=
\op{S}_{p_1}^{-}\,\op{S}_{p_1}^{-}\,\cdots\op{S}_{p_l}^{-}\,
\ket{\Omega}
\end{eqnarray}
%--------------------------------------------------------
for $p_{\nu}\in\widehat{G}, \nu=1,\dots,l$ and 
$l=0,\dots,2Ns$,
$\ket{\Omega}$ being the magnon vacuum state as given in
\eqref{E-3-7}.
For $l>2Ns$ $\ket{\Omega_{\vek{p}}}$ will be the zero vector.
Since the $\op{S}_{p_{\nu}}^{-}$ commute, a
unique representation of \fmref{E-4-9} may be achieved by
postulating $p_1\le p_2\le \dots \le p_l$.

One-magnon states $\ket{\Omega_{p}}$ are orthogonal
%--------------------------------------------------------
\begin{eqnarray}
\label{E-4-10}
\braket{\Omega_{p}}{\Omega_{q}}
=
2 s \delta_{p,q}
\ ,
\end{eqnarray}
%--------------------------------------------------------
which follows from \fmref{E-4-7-2} and 
%--------------------------------------------------------
\begin{eqnarray}
\label{E-4-11}
\op{S}_{p}^{3}\,\ket{\Omega}
=
\sqrt{N}\, s\, \delta_{p,0}\,\ket{\Omega}
\ .
\end{eqnarray}
%--------------------------------------------------------
Moreover, the one-magnon states are already eigenvectors of
$\op{H}$ and $\op{T}$
%--------------------------------------------------------
\begin{eqnarray}
\label{E-4-12}
\op{H}\ket{\Omega_{p}}
&=& 
E_p\,\ket{\Omega_{p}}
\ ,\quad 
E_p=2 s \tilde{J}_p + s (N s -2) \tilde{J}_0
\ ,
\\
\op{T}\ket{\Omega_{p}}
&=& 
e^{-i p}\,\ket{\Omega_{p}}
\ .
\end{eqnarray}
%--------------------------------------------------------
The last relation may be derived from 
%--------------------------------------------------------
\begin{eqnarray}
\label{E-4-13}
\op{T}^x\,\op{S}_p^{\pm}
=
e^{-i p x}\,\op{S}_p^{\pm}\,\op{T}^x
\quad
\forall x\in G,\ \forall p \in\widehat{G}\ .
\end{eqnarray}
%--------------------------------------------------------
For $l>1$ the $l$-magnon states are in general no longer
orthogonal or linearly independent. But they span the total
Hilbert space ${\mathcal H}$ and they behave nicely with respect
to the ladder operator $\op{S}^{-}$, namely
%--------------------------------------------------------
\begin{eqnarray}
\label{E-4-14}
\op{S}^{-}\, \ket{\Omega_{p_1,\dots,p_l}}
=
\sqrt{N}\, \ket{\Omega_{0,p_1,\dots,p_l}}
\ ,
\end{eqnarray}
%--------------------------------------------------------
and they span the subspaces ${\mathcal H}(M,p)$,
which follows from
%--------------------------------------------------------
\begin{eqnarray}
\label{E-4-15}
\op{T}\, \ket{\Omega_{p_1,\dots,p_l}}
&=&
\exp\left\{-i\sum_{\nu=1}^l p_{\nu}   \right\}
\, \ket{\Omega_{p_1,\dots,p_l}}
\\
\op{S}_{0}^{3}\,\ket{\Omega_{p_1,\dots,p_l}}
&=&
(N s - l)\,\ket{\Omega_{p_1,\dots,p_l}}
\ .
\end{eqnarray}
%--------------------------------------------------------
Let the magnon states $\ket{\Omega_{\vek{p}}}$ be ordered as
follows
%--------------------------------------------------------
\begin{eqnarray}
\label{E-4-16}
&&\ket{\Omega}
\\
&&\ket{\Omega_{0}}, \ket{\Omega_{1}}, \dots,
\ket{\Omega_{N-1}}
\nonumber \\
&&\ket{\Omega_{00}}, \ket{\Omega_{11}}, \dots,
\ket{\Omega_{N-1,N-1}}
\nonumber \\
&&\quad\vdots
\nonumber
\end{eqnarray}
%--------------------------------------------------------
where each row is ordered lexicographically.  We will call the
basis of ${\mathcal H}$ generated by applying the well-known
Gram-Schmidt orthonormalization procedure to the sequence
\fmref{E-4-16} ``magnon basis". Note that the Gram-Schmidt
procedure will sometimes produce zero vectors, because the
$\ket{\Omega_{\vek{p}}}$ are linearly dependent.  The main
result of this section is then the following proposition.

{\bf Proposition:} The magnon basis contains a subbasis for each
subspace ${\mathcal H}(S,M,p)$, $S=Ns, Ns-1,\dots, |M|;\
p\in\widehat{G}$.

{\bf Proof:}
Since the $\ket{\Omega_{\vek{p}}}$ span the subspaces
${\mathcal H}(M,p)$ and different ${\mathcal H}(M,p)$ are
orthogonal, the magnon basis spans ${\mathcal H}(M,p)$, too. The
subspace with maximal $S=Ns$ is spanned by the magnon states
%--------------------------------------------------------
\begin{eqnarray}
\label{E-4-17}
\ket{\Omega}, \ket{\Omega_{0}}, \ket{\Omega_{00}}, \dots
\ket{\Omega_{(2 N s) \mbox{zeros}}}
\ ,
\end{eqnarray}
%--------------------------------------------------------
which also occur in the magnon basis up to normalization. These
vectors also span the one-dimensional spaces ${\mathcal
H}(S,M,p)$ with $S=Ns; M=-S,\dots,S; p=0$. Hence the proposition
holds for $S=Ns$. We then may proceed by induction. Under the
assumption that the proposition holds for all $S>S_0$, we have
to show that it holds for $S=S_0$.  Consider the subspaces
${\mathcal H}(M)$ with $M=S_0$. According to \fmref{E-3-13} we
have
%--------------------------------------------------------
\begin{eqnarray}
\label{E-4-18} 
{\mathcal H}(S_0)
=
\op{S}^{-}\,{\mathcal H}(S_0+1)
\oplus
{\mathcal H}(S_0,S_0)
\ .
\end{eqnarray}
%--------------------------------------------------------
Hence ${\mathcal H}(S_0,S_0)$ is spanned by vectors from the
magnon basis and the same holds for every ${\mathcal H}(S_0,S_0,p),
p\in\widehat{G}$. For the other subspaces ${\mathcal
H}(S_0,M,p),M<S_0,$ we have
%--------------------------------------------------------
\begin{eqnarray}
\label{E-4-19} 
{\mathcal H}(S_0,M,p)
=
\left(\op{S}^{-}\right)^{S_0-M}\,{\mathcal H}(S_0,S_0,p)
\end{eqnarray}
%--------------------------------------------------------
and we know that the restrictions
%--------------------------------------------------------
\begin{eqnarray}
\label{E-4-20} 
\op{S}^{-}:\quad
{\mathcal H}(M+1)\longrightarrow
{\mathcal H}(M)\ ,\quad M\ge 0,
\end{eqnarray}
%--------------------------------------------------------
are isometries up to a factor $\sqrt{S(S+1)-M(M+1)}$. Hence
$\op{S}^{-}$ maps vectors from the magnon basis onto vectors
from the magnon basis (up to a factor), which concludes the
proof. \hfill $\blacksquare$

We are left with two tasks. First, in order to calculate the
magnon basis, we need a formula for the inner product between
magnon states. We have no explicit formula, but using the
commutation relations \fmref{E-4-7-1} and \fmref{E-4-7-2} we derived
a recursion relation, which is sufficient for computer algebraic
calculations. Similarly, we can express the matrix elements of
the Hamiltonian between two magnon states by finite sums
containing only inner products of magnon states, which completes
the second task. The results are
%--------------------------------------------------------
\begin{eqnarray}
\label{E-4-21}
\braket{\Omega_{p_1,\dots,p_l}}{\Omega_{q_1,\dots,q_l}}
&=&
2 s \sum_{\nu=1}^l\,
\delta_{q_{\nu},p_1}\,\braket{\Omega_{p_2,\dots,p_l}}{\Omega_{q_1,\dots,\check{q}_{\nu},\dots,q_l}}
\\
&&
-
\frac{2}{N}
\sum_{1\le \nu < \mu \le l}\,
\braket{\Omega_{p_2,\dots,p_l}}{\Omega_{q_1,\dots,\check{q}_{\nu},\dots,(q_{\mu}\mapsto(q_{\mu}+q_{\nu}-p_1)),\dots,q_l}}
\nonumber
\ ,
\end{eqnarray}
%--------------------------------------------------------
where the symbol $\check{q}_{\nu}$ denotes deletion of the
index $q_{\nu}$ and $(q_{\mu}\mapsto(q_{\mu}+q_{\nu}-p_1))$
denotes replacement of the index $q_{\mu}$ by the given
expression which has to be understood modulo $N$.
Further
%--------------------------------------------------------
\begin{eqnarray}
\label{E-4-22}
&&
\bra{\Omega_{p_1,\dots,p_l}}\op{H}\ket{\Omega_{q_1,\dots,q_l}}
\\
&=&
\sum_{p\in\widehat{G}}\;
\tilde{J}_p\,
\braket{\Omega_{p,p_1,\dots,p_l}}{\Omega_{p,q_1,\dots,q_l}}
+
\tilde{J}_0\,\braket{\Omega_{p_1,\dots,p_l}}{\Omega_{q_1,\dots,q_l}}
\left(
s^2 N - 2 s l + \frac{l^2}{N}
\right)
\nonumber \\
&&+
\sum_{p\in\widehat{G},p\ne 0}\;
\tilde{J}_p\,
\frac{1}{N}
\sum_{\nu,\mu=1}^l\,
\braket{\Omega_{p_1,\dots,(q_{\nu}\mapsto(q_{\nu}+p)),\dots,p_l}}
{\Omega_{q_1,\dots,(q_{\mu}\mapsto(q_{\mu}+p)),\dots,q_l}}
\nonumber
\ .
\end{eqnarray}
%--------------------------------------------------------
Now we have all what is required in order to write a program
\cite{HJS99} which reduces the diagonalization of $\op{H}$ to
the diagonalization of the submatrices of $\op{H}$ with respect
to the magnon basis of the relevant subspaces ${\mathcal
H}(S,M,p)$. The final diagonalization can be done numerically or
in the few cases, where the relevant dimension is less or equal
4, analytically. Table \xref{T-3-1} shows in bold face for which
cases analytical solution will be possible. Two examples for
eigenvalues and multiplicities are given in the next
section. The complete routine, which also presents the
eigenvectors, may be downloaded from the internet
\cite{HJS99}. It may be possible to exactly solve other cases if
the discrete symmetries mentioned in the introduction are taken
into account.  The cases $N=2,3,4$ can, in principle, also be solved
by Clebsch-Gordan decomposition of the Hilbert space.

\section{Examples}
\label{sec-5}

\subsection{Relevant dimension of spin rings}

%-----------------------------------------------------------------------
\begin{table}[t]
\begin{center}
\begin{tabular}{|cc||ccc|ccc|ccc|ccc|ccc|ccc|}
\hline
&&\multicolumn{18}{c|}{$k$}\\
&& &0&&&1&&&2&&&3&&&4&&&5&\\
\hline
&S& 3&1&0& 2&1&0& 2&1&& 2&1&0& 2&1&& 2&1&0\\
\hline\hline
\multirow{7}{5mm}{$M$}
& 3& 1&&& &&& &&& &&& &&& &&\\
& 2& 1&&& 1&&& 1&&& 1&&& 1&&& 1&&\\
& 1& 1&2&& 1&1&& 1&2&& 1&1&& 1&2&& 1&1&\\
& 0& 1&2&1& 1&1&1& 1&2&& 1&1&2& 1&2&& 1&1&1\\
&-1& 1&2&& 1&1&& 1&2&& 1&1&& 1&2&& 1&1&\\
&-2& 1&&& 1&&& 1&&& 1&&& 1&&& 1&&\\
&-3& 1&&& &&& &&& &&& &&& &&\\
\hline
\end{tabular}
\vspace*{5mm}
\end{center}
\caption{Dimensions of the subspaces ${\mathcal H}(S,M,p(k))$ for
$N=6$, $s=\half$. Each column can be created by applying the
ladder operator \fmref{E-3-11} yielding $(2S+1)$-dimensional
subspaces. One realizes a discrete symmetry between columns for $k$
and $6-k$.}\label{T-5-3} 
\end{table}
%----------------------------------------------------------------------- 

%-----------------------------------------------------------------------
\begin{table}[ht]
\begin{center}
\begin{tabular}{|cc||r|r|r|r|r|r|r|r|r|}
\hline
&&\multicolumn{9}{c|}{$N$}\\
&& 2&3&4&5&6&7&8&9&10\\
\hline\hline
\multirow{10}{5mm}{$s$}
&$\frac{1}{2}$& { 1} & { 2} & { 3} & 5 & 9 & 14 & 28 & 48 & 90 \\
&$\frac{1}{2}$& {\bf 1} & {\bf 1} & {\bf 1} & {\bf 1} & {\bf 2} & {\bf 2}  & {\bf 4}  & 6  & 10 \\
\cline{2-11}
&$1$& { 1} & { 3} & 6 & 15 & 40 & 105 & 280 & 750 & 2025 \\
&$1$& {\bf 1} & {\bf 1} & {\bf 2} & {\bf 3}  & 8  & 15  & 37  & 84  & 207  \\
\cline{2-11}
&$\frac{3}{2}$& { 1} & { 4} & 11 & 36 & 120 & 426 & 1505 & 5300 & 19425 \\
&$\frac{3}{2}$& {\bf 1} & {\bf 2} & {\bf 4}  & 8  & 23  & 61  & 192  & 590  & 1956 \\
\cline{2-11}
&$2$& { 1} & 5 & 17 & 70 & 295 & 1260 & 5620 & 25200 & 113706 \\
&$2$& {\bf 1} & {\bf 2} & 5  & 14 & 53  & 180  & 712  & 2800  & 11403  \\
\cline{2-11}
&$\frac{5}{2}$& { 1} & 6 & 24 & 120 & 609 & 3150 & 16576 & 88900 & 484155 \\
&$\frac{5}{2}$& {\bf 1} & {\bf 2} & 7  & 24  & 105 & 450  & 2085  & 9884  & 48483  \\
\hline
\end{tabular}
\vspace*{5mm}
\end{center}
\caption{Relevant dimension assuming only invariance with
respect to rotations (upper rows) and assuming also invariance
with respect to cyclic shifts (lower rows).}\label{T-3-1}
\end{table}
%----------------------------------------------------------------------- 

In order to illustrate the above considerations let us discuss
the example $N=6$, $s=\half$, which has been solved by
D.~Kouzoudis \cite{Kou98}. The decomposition \fmref{E-3-2}
yields the following relation between the dimensions of the
involved Hilbert spaces
%--------------------------------------------------------
\begin{eqnarray}
\label{E-5-1}
\mbox{dim}\left({\mathcal H}\right)
=
2^6
=
1 + 6 + 15 + 20 + 15 + 6 + 1
=
\sum_{M=-\SMax}^{\SMax}\!\!\mbox{dim}\left({\mathcal H}(M)\right)
\ .
\end{eqnarray}
%--------------------------------------------------------
Obviously, $\ket{\Omega}=\ket{++++++}$ spans the
one-dimensional space ${\mathcal H}(M)$ with $M=\SMax=Ns=3$,
which is an epicycle of dimension $D=1$. ${\mathcal H}(M=2)$
consists of the proper cycle generated e.g. by $\ket{-+++++}$
and hence is six-dimensional. It is spanned by the six
eigenstates of $\op{T}$ and $\op{H}$ given in \eqref{E-3-22},
which are one-magnon states.
The next space ${\mathcal H}(M=1)$ consists of two proper cycles
generated e.g. by $\ket{--++++}$ and $\ket{-+-+++}$,
respectively, and one
three-di\-men\-sional epicycle generated by $\ket{-++-++}$. Hence 
$\mbox{dim}\left({\mathcal H}(M=1)\right)=6+6+3=15$. Note that
the last mentioned epicycle corresponds to the proper cycle
generated by $\ket{-++}$ with $N=3$, $M=\half$. The largest
space ${\mathcal H}(M=0)$ is spanned by three proper cycles
generated by $\ket{--+-++}$, $\ket{++-+--}$, $\ket{+++---}$
and one two-dimensional epicycle generated by
$\ket{+-+-+-}$. Consequently, 
$\mbox{dim}\left({\mathcal H}(M=0)\right)=6+6+6+2=20$. The
remaining spaces ${\mathcal H}(M)$ with $M<0$ have the same
cycle structure as ${\mathcal H}(|M|)$ due to the symmetry
$+\leftrightarrow -$.

The dimensions of the subspaces ${\mathcal H}(S,M,p(k))$ are given
in table \xref{T-5-3}.
Note that in accordance with the rule of thumb \fmref{E-3-18}
the numbers $\mbox{dim}\left({\mathcal H}(\cdot,M,p)\right)$ are
almost uniformly distributed with respect to $p$. The deviations
from the uniform distribution for $|M|=0,1$ can be explained by
the extra eigenvalues of $\op{T}$ produced by the mentioned
epicycles of dimensions three and two.

The next table \xref{T-3-1} shows the relevant dimensions for a
wide variety of $N$ and $s$. The ratio of upper-row to lower-row
entries is roughly $N$ in accordance with the rule of thumb,
\eqref{E-3-18}. Dimensions less than five are given in bold,
these cases can be solved analytically. Two examples are
presented in the next subsection.

%%%%%%%%%%%%%%%%%%%%%%%%%%%%%%%%%%%%%%%%%%%%%
\subsection{Exactly solvable systems}

\begin{figure}[t]
\begin{center}
\epsfig{file=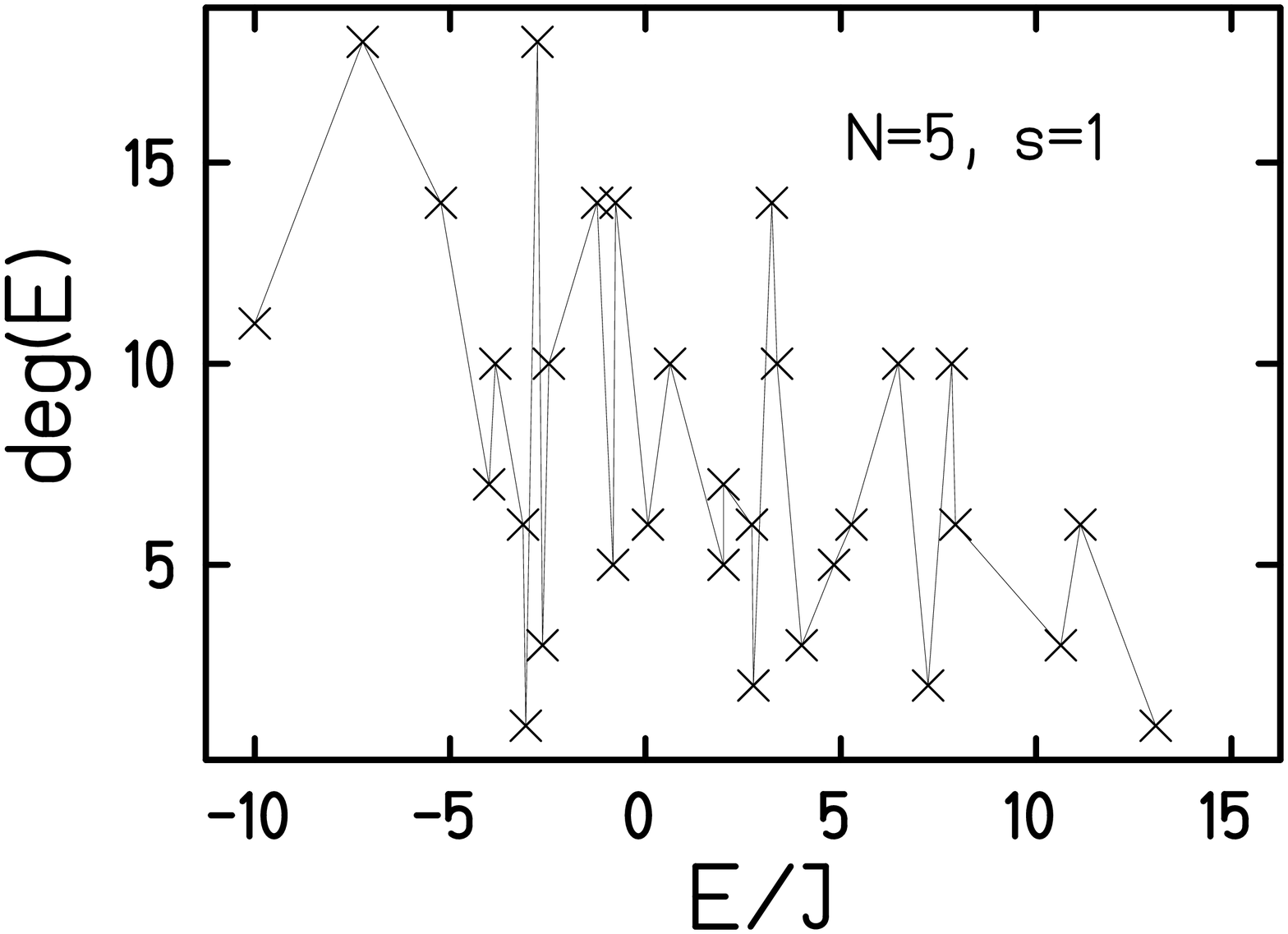,width=60mm}
$\qquad$
\epsfig{file=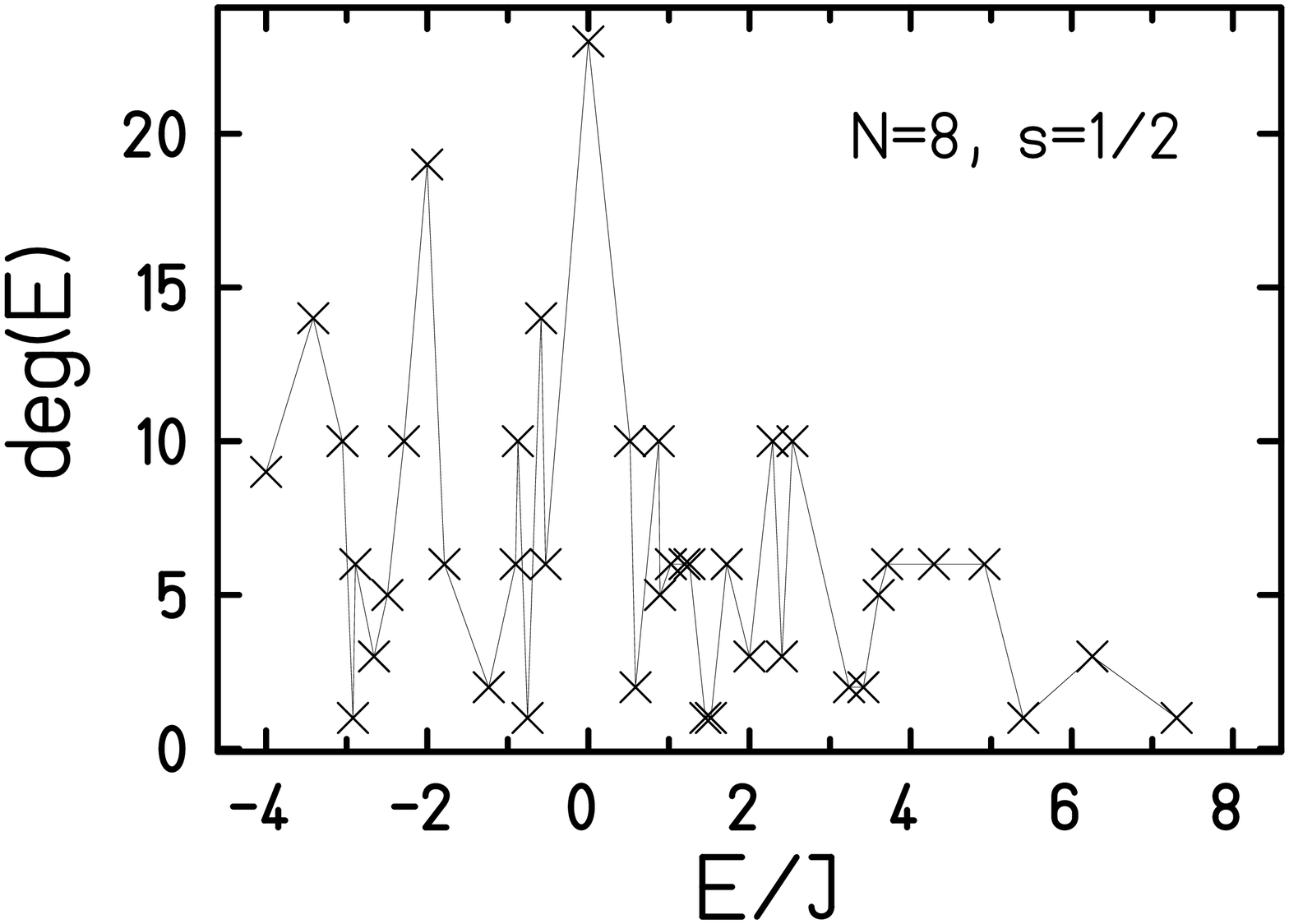,width=60mm}
\caption{Eigenvalues and degeneracies for the Heisenberg
Hamiltonian \eqref{E-4-1} with $N=5, s=1$ (l.h.s.) and
$N=8,s=\half$ (r.h.s.). The lines are drawn as a guide for the
eye.}
\label{F-5-1}
\end{center} 
\end{figure}

Analytical solutions are for instance possible in the cases
$N=5,s=1$ and $N=8,s=\half$, which to our knowledge have not yet
been published. Table \xref{T-5-1} contains the exact
eigenvalues and multiplicities for the Heisenberg Hamiltonian
\eqref{E-4-1} with nearest neighbour interaction and $N=5,s=1$
and \figref{F-5-1} shows the same information graphically, but
also for $N=8,s=\half$. The analytical, but rather lengthy
expressions for the latter case may be evaluated using a
Mathematica$^{\circledR}$ script we provide
\cite{HJS99}. Here we would like to present only the eigenvalue
which in the antiferromagnetic case corresponds to the ground
state energy
%--------------------------------------------------------
\begin{eqnarray}
\label{E-5-3}
E_0/J
=
\frac{4}{3}
\left\{
2 + \sqrt{13}\cos\left[\frac{1}{3}\mbox{arctan}\left(\frac{3\sqrt{3}}{5}\right) \right]
\right\}
\ \mbox{for}\ N=8,s=\half
\ .
\end{eqnarray}
%--------------------------------------------------------

%-----------------------------------------------------------------------
\begin{table}[t]
\begin{flushleft}
\begin{tabular}{|c||c|c|c|c|c|c|c|c|c|}
\hline
E/J &
-10 & -4 & 2&$2\pm 2\sqrt{2}$&4&$4\pm 2\sqrt{11}$&
$5\pm 2\sqrt{65}$&$5\pm \sqrt{5}$&$-5\pm \sqrt{5}$\\
\hline
deg(E)&
11 &   7 &12&15              &3&3                &
1                &2              &18\\
\hline
\end{tabular}
\begin{tabular}{|c||c|c|c|c|}
\hline
E/J &
$-3\pm \sqrt{5}$&$1\pm \sqrt{5}$&$2\pm \sqrt{5}\pm \sqrt{13}$&
$\frac{2}{3}\left(6\pm\left[\sqrt{5}+2\sqrt{23}\cos(\alpha)\right]\right)$\\
\hline
deg(E)&
14              &14             &10                          &
6\\
\hline
\end{tabular}
\begin{tabular}{|c||c|}
\hline
E/J &
$\frac{2}{3}\left(6\pm\left[\sqrt{5}-\sqrt{23}\cos(\alpha)\right]
\pm\sqrt{69}\sin(\alpha)\right)$\\
\hline
deg(E)&
6\\
\hline
\end{tabular}
\vspace*{5mm}
\end{flushleft}
\caption{Eigenvalues and degeneracies for the Heisenberg
Hamiltonian \eqref{E-4-1} with nearest neighbour interaction and
$N=5, s=1$. The value of $\alpha$ is
$\alpha=\frac{1}{3}\mbox{arctan}\left[\sqrt[3]{\frac{1338}{5}}/5\right]$}\label{T-5-1}
\end{table}
%----------------------------------------------------------------------- 

%===================    figure   =================================
%===================    figure   =================================

%===================    figure   =================================
\begin{figure}[t]
\begin{center}
\epsfig{file=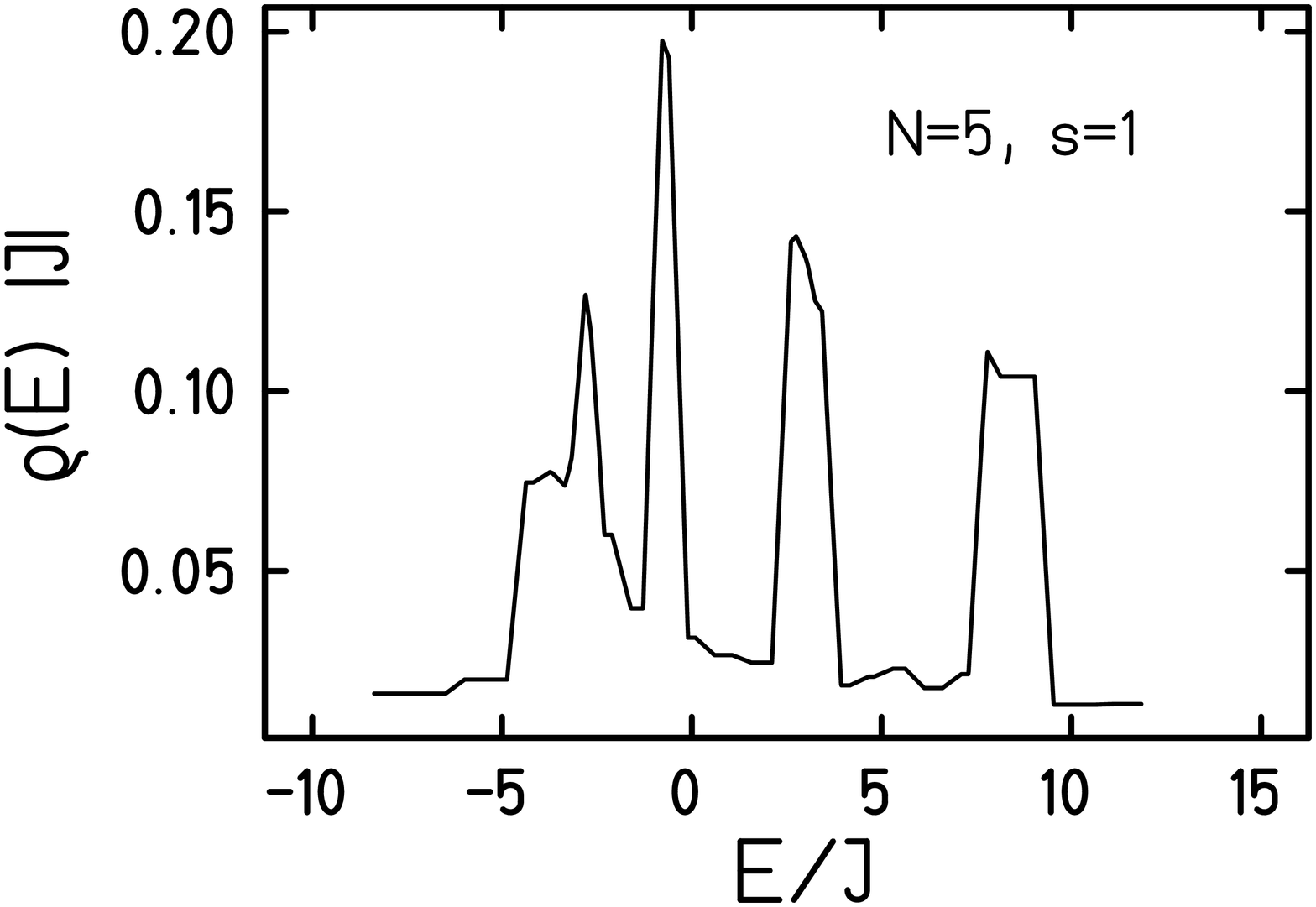,width=60mm}
$\qquad$
\epsfig{file=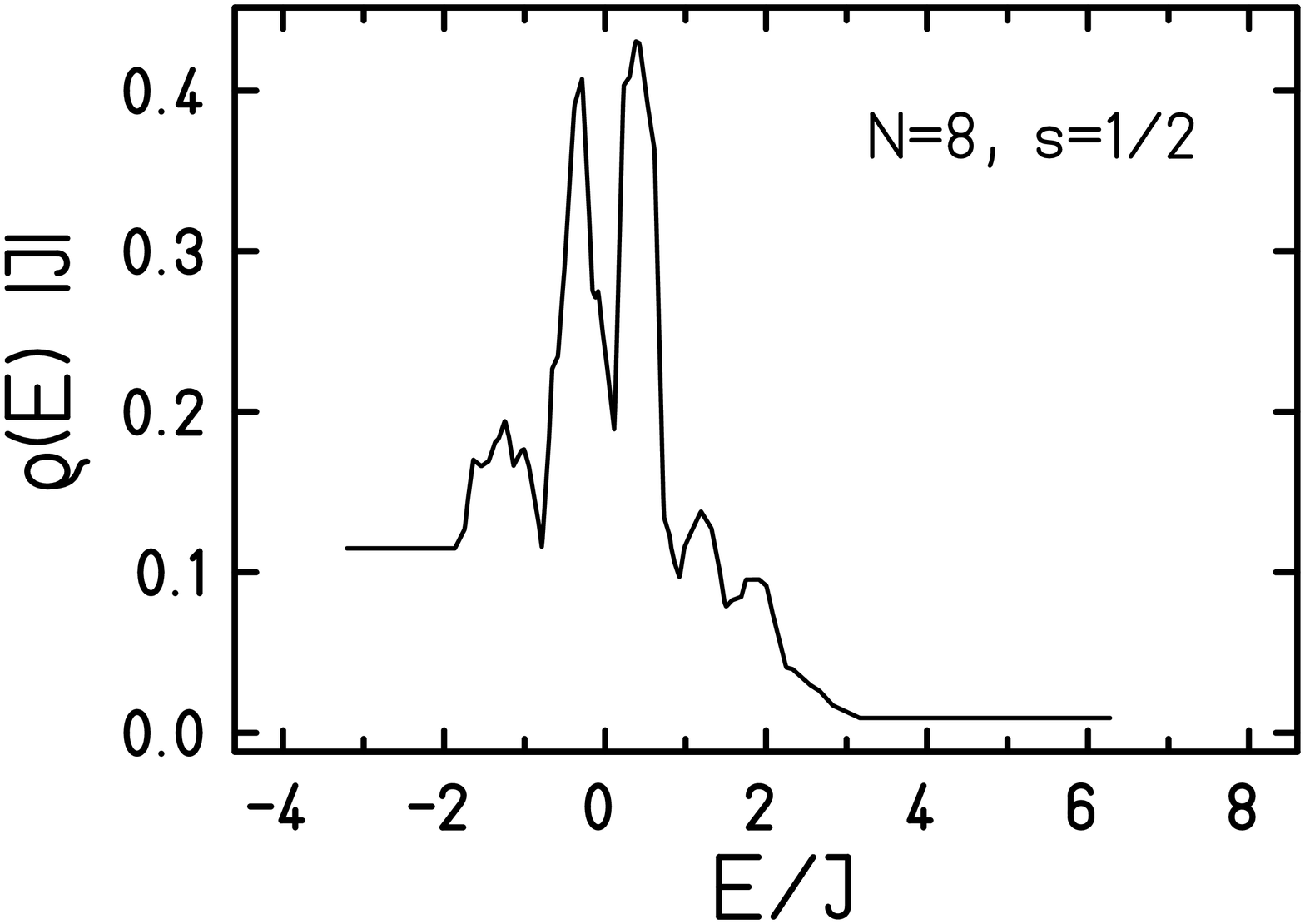,width=60mm}
\caption{Smoothened density of states $\rho(E)$ for the Heisenberg
hamiltonian \eqref{E-4-1} with $N=5, s=1$ (l.h.s.) and
$N=8,s=\half$ (r.h.s.).}
\label{F-5-3}
\end{center} 
\end{figure} 
%===================    figure   =================================

In order to clarify the structure of the energy spectrum we
calculate the smoothened density of states. To this end the
degeneracy of an energy eigenvalue $E_n$ is first divided by the
mean distance to the neighbouring eigenvalues, then the
resulting function is linearly interpolated and convoluted with
a characteristic function of width 1. This density of states
$\rho(E)$, which is displayed for the Heisenberg Hamiltonian
\eqref{E-4-1} in \figref{F-5-3} for $N=5, s=1$ (l.h.s.) and
$N=8,s=\half$ (r.h.s.), shows an interesting band structure.

%%%%%%%%%%%%%%%%%%%%%%%%%%%%%%%%%%%%%%%%%%%%%
\subsection{Bethe-Hulth\'{e}n}

%===================    figure   =================================
\begin{figure}[t]
\begin{center}
\epsfig{file=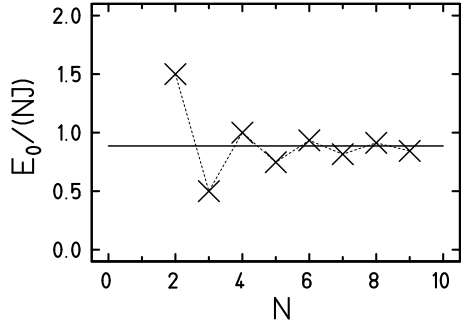,width=60mm}
\caption{Comparison of analytical ($N\le 8$) and numerical
($N>8$) ground state energies (symbols) for antiferromagnetic coupling
with the large $N$ limit (solid line) of Bethe and Hulth\'{e}n
\cite{Bet31,Hul38}.}
\label{F-5-2}
\end{center} 
\end{figure} 
%===================    figure   =================================

The calculated eigenvalues, Table \xref{T-5-1} and Figure
\xref{F-5-1}, correspond to the ferromagnetic case for $J>0$ and to
the antiferromagnetic case for $J<0$. Thus our data also contain
the antiferromagnetic ground state energy $E_0(N)$
and we can compare $E_0(N)/(NJ)$ with the exact limit
%--------------------------------------------------------
\begin{eqnarray}
\label{E-5-2}
\lim_{N\rightarrow\infty}\,
\frac{E_0(N)}{NJ}
=
2\ln(2) - \half
\end{eqnarray}
%--------------------------------------------------------
known as the Bethe-Hulth\'{e}n formula \cite{Yos96,Bet31,Hul38}. In
\figref{F-5-2} analytical ($N\le 8$) and numerical
($N>8$) ground state energies are displayed by symbols and the
large $N$ limit by a solid line. It turns out that the limit is
approached from above for even $N$ and from below for odd $N$
and that the approach is faster for even values of $N$. This
shows the effect of spin frustration. 

%\clearpage
%%%%%%%%%%%%%%%%%%%%%%%%%%%%%%%%%%%%%%%%%%%%%%%%%%%%%%%%%%%%%%%%%%%%%%%%
%
%{\bf Acknowledgments}\\[5mm]
\section*{Acknowledgments}
The authors would like to thank
M.~Luban and D.~Mentrup for helpful discussions.

%
%%%%%%%%%%%%%%%%%%%%%%%%%%%%%%%%%%%%%%%%%%%%%%%%%%%%%%%%%%%%%%%%%%%%%%%%

\end{document}